\documentclass[referee]{raa}
\usepackage{graphicx,times}
\usepackage{natbib}
\usepackage{amssymb,amsmath}
\bibpunct{(}{)}{;}{a}{}{,}

\usepackage[a4paper=true,dvipdfm=true,pagebackref=true]{hyperref}
\hypersetup{pdftitle = The title of my PDF, pdfauthor = My name, pdfsubject= The subject, pdfkeywords = keyword1 keyword2 keyword3}
\hypersetup{colorlinks = true, linkcolor = green, anchorcolor = red, citecolor = blue, filecolor = red, pagecolor = red, urlcolor = red}

\begin{document}

   \title{Pulsar timing residuals due to  individual non-evolving gravitational
 wave   sources$^*$
\footnotetext{\small $*$ Supported by National
Natural Science Foundation of China}
}

 \volnopage{ {\bf 2012} Vol.\ {\bf X} No. {\bf XX}, 000--000}
   \setcounter{page}{1}

   \author{Ming-Lei Tong\inst{1,2}, Cheng-Shi Zhao\inst{1,2}, Bao-Rong Yan\inst{1,3}, Ting-Gao Yang,
      \inst{1,2},  Yu-Ping Gao\inst{1,2}
   }

   \institute{ National Time Service Center, Chinese Academy of Sciences, Xi'an 710600;
    {\it mltong@ntsc.ac.cn}\\
        \and
             Key Laboratory of Time and Frequency Primary Standards,
Chinese Academy of Sciences, Xi'an 710600\\
	\and
	  Key Laboratory of Precision Navigation and Timing Technology,
Chinese Academy of Sciences, Xi'an 710600\\
\vs \no
   {\small Received xxx; accepted xxx}
}

\abstract{The pulsar timing residuals induced by gravitational waves from  non-evolving single binary
 sources are affected by many parameters related to the relative positions of the pulsar and
 the gravitational wave sources.
  We will fully  analyze the  effects due to different parameters one by one.  The standard deviations of the timing residuals will be calculated with a variable parameter fixing a set of other  parameters.    The orbits of the binary sources will be generally assumed to be elliptical. The influences of  different eccentricities on the pulsar timing residuals will also studied in detail. We find that effects
  of the related parameters are quite different, and some of them present  certain regularities.
}

   \authorrunning{M.-L. Tong et al. }            
   \titlerunning{Pulsar timing residuals due to  individual non-evolving gravitational
 wave   sources}  
   \maketitle

%
\section{Introduction}           
\label{sect:intro}

Gravitational waves (GWs) are believed to exist in the universe according to  general relativity. The detection of GWs has been a very
interesting field which has attracted  much attention of many scientists.
Even though the indirect evidence of GW emission was provided by the observations of the binary pulsar
B1913+16 (\citealt{Hulse+Taylor+1974}), there is no direct GW signals to be detected so far. However,  many methods of direct detection of GWs have
 been proposed and tried for a long time. Various GW detectors were constructed or proposed for different frequencies, including the ground-based interferometers, such as Advanced LIGO \footnote{http://www.ligo.caltech.edu/advLIGO} and KAGRA (\citealt{Somiya+2012}) aiming at $10^2-10^3$ Hz; the  space-based interferometers, such as eLISA \footnote{http://elisa-ngo.org}
 and ASTROD-GW (\citealt{Ni+2013})  aiming at $10^{-5}-10^{-1}$ Hz;  pulsar timing arrays (\citealt{Sazhin+1978}; \citealt{Detweiler+1979}; \citealt{Foster+Backer+1990}; \citealt{Jenet+etal+2005}; \citealt{Hobbs+etal+2009}) aiming at $10^{-9}-10^{-7}$ Hz,  waveguide (\citealt{Cruise+2000}; \citealt{Tong+Zhang+2008}) aiming at $10^{-6}-10^{-8}$ Hz, Gaussian beam (\citealt{Li+etal+2003}; \citealt{Tong+etal+2008}) aiming at GHz, and even the anisotropies
and polarizations of the cosmic microwave background radiation (\citealt{Zaldarriaga+Seljak+1997}; \citealt{Kamionkowski+etal+1997}) aiming at $10^{-18}$ Hz.

 With the improvement of radio telescopes,  more and more pulsars are founded, and, moreover, the measurement technique is more and more precise, which  ensure
 pulsar timing arrays (PTAs) be  powerful  in detecting  GWs directly.  The times-of-arrival (TOAs) of the pulses radiated from pulsars will be fluctuated as  GWs
 passing through the path between the pulsars and the earth. As shown in \citealt{Hellings+Downs+1983}, a stochastic GW background can be detected by searching for correlations in the timing residuals of an array of millisecond pulsars spread over the sky. On the other hand, single sources of GWs are also important in the observations of pulsar timing array (\citealt{Lee+etal+2011})  or an individual pulsar \citealt{Jenet+etal+2004}.
 Currently,
 there are several  PTAs running,
such as the Parkes Pulsar Timing Array (PPTA) (\citealt{Manchester+etal+2013}),
European Pulsar Timing Array (EPTA) (\citealt{van Haasteren+etal+2011}),
the North American Nanohertz Observatory for Gravitational Waves (NANOGrav) (\citealt{Demorest+etal+2013}),
and the International Pulsar Timing Array (IPTA) (\citealt{Hobbs+etal+2010}) formed by the aforesaid three PTAs.
Moreover,  much more sensitive Five-hundred-meter Aperture Spherical Radio Telescope (FAST) (\citealt{Nan+etal+2011}) and  Square Kilometre Array (SKA) \footnote{http://www.skatelescope.org}
are also under planning.

There are chiefly three kinds of GW sources:
continuous sources (\citealt{Peters+1964}),  instantaneous  sources (\citealt{Thoren+Braginskii+1976}) and the stochastic gravitational wave background (\citealt{Grishchuk+1975}; \citealt{Starobinsky+1979}; \citealt{Zhang+etal+2005}; \citealt{Tong+2013}; \citealt{Jaffe+Backer+2003}; \citealt{Damour+Cilenkin+2005}).
In the range of $\sim10^{-9}-10^{-7}$ Hz,  the
  major targets are GWs generated by  supermassive black hole binaries (SMBHBs) (\citealt{Jaffe+Backer+2003}; \citealt{Sesana+Vecchio+2010}).
  As analyzed in \citealt{Lee+etal+2011}, a PTA is sensitive to  the nano-hertz GWs from SMBHB systems with masses of $\sim 10^8-10^{10}M_\odot$ less than $10^5-10^6$ years
 before the final merger. Binaries with more than $\sim10^3-10^4$ years before merger can be
 treated as  non-evolving GW sources. The non-evolving SMBHBs are believed to be the dominant population, since they have lower masses and longer rest lifetimes.
 The pulsar timing residuals, defined as the difference between the observational times-of-arrival and those predicted by the pulsar timing model, provide us very important information. For example, one can fit the observational data using the least square method to
 obtain the rotational frequency and its first derivative. As for GWs, the timing residuals of a pulsar timing array can be used to extract GW signals from various noises due to the correlations of the signals. On the other hand, GWs give rise to additional timing residuals, which will affect the precision of the  pulsar timing standard. Hence, the studies of the timing residuals induced by GWs are important.  In this paper, we fully  analyze   the timing residuals of an individual pulsar induced by single  non-evolving GW sources of SMBHBs localized in various positions in the frame of  celestial reference.    They are related to many parameters.  We will discuss the effects of all the parameters  on the timing residuals and their standard deviations one by one. Following the discussions in Refs. (\citealt{Wahlquist+1987}; \citealt{Tong+etal+2013a}),  we assume generally the orbits of the SMBHBs are elliptical, i.e., the eccentricities will be nonzero.   For example,  one of the best-known candidates for a SMBHB system emitting
GWs with frequency detectable by pulsar timing is in the blazar OJ287 (\citealt{Sillanpaa+etal+1996}), with an orbital
eccentricity $e\sim0.7$ (\citealt{Lehto+Valtonen+1997}; \citealt{Zhang+etal+2013}). Moreover, if we consider the existence of the circumbinary
gaseous discs in which the SMBHBs black hole binaries embedded, the binary orbits are usually
eccentric (\citealt{Rodig+etal+2011}). However, we will not consider the angular momentum transfer between the SMBHBs and their self-gravitating discs.    The effect of different values of the eccentricities on the
 standard deviations of the timing residuals  will also be studied. 
  
  In section 2, we show the analytical solution of GWs from an SMBHB with a general
elliptical orbit. Section 3 describes how a singe GW induces pulsar timing residuals. Various
effects due to different parameters on the standard deviations of the pulsar timing residuals
will be analyzed in section 4. Section 5 provides some conclusions and discussions.   Throughout this paper, we use units in which $c=G=1$.

\section{The analytic solution of GWs from a SMBHB with an elliptical orbit}  

First of all, we  simply describe the derivations of the GW solution from a binary star system with an elliptical orbit  following \citealt{Wahlquist+1987}.
 The usual equation for the
 relative orbit ellipse is given by
 \begin{equation}
 r=\frac{a(1-e^2)}{1+e \cos(\theta-\theta_p)},
 \end{equation}
 where $r$ is the relative separation of the binary components, $a$ is the semi-major axis, and $\theta_p$ is the value
 of $\theta$ at the periastron. Since different values of $\theta_p$  correspond to different choices of the initial time,   we set concretely $\theta_p=180^\circ$ without losing generality in the following.  The orbital period of the binary system  is
 \begin{equation}
 P=\left(\frac{4\pi^2a^3}{M}\right)^{1/2},
 \end{equation}
 where  $M$ is the total mass of the binary system.
 According to the differential equation for the Keplerian motion, one get (\citealt{Wahlquist+1987})
\begin{equation}
\dot{\theta}=(2\pi/P)(1-e^2)^{-3/2}[1+e\cos{(\theta-\theta_p)}]^2.
\end{equation}
Integrating the above equation, one has the relation between $t$ and $\theta$, which is
shown in Figure 1 in \citealt{Tong+etal+2013a}.

For observations, one needs to  obtain the solution of GWs described in the frame  where the origin locates at the Solar System Barycenter (SSB). Thus, first of all, let us construct the solar barycenter celestial reference frame (BCRF) whose origin is SSB.
 Let $\{\hat{i},\hat{j},\hat{k}\}$ be the base vectors of the BCRF.
Then the unit vector of a GW source is
\begin{equation}
\hat{d}=\cos\delta(\cos\alpha\,\hat{i}+\sin\alpha\, \hat{j})+\sin\delta\,\hat{k},
\end{equation}
where $\alpha$ and $\delta$ are the right ascension and declination of the binary source, respectively. Define orthonormal
vectors on the celestial sphere by (\citealt{Wahlquist+1987})
\begin{equation}\label{alpha}
\hat{\alpha}\equiv-\sin\alpha\,\hat{i}+\cos\alpha\,\hat{j},
\end{equation}
\begin{equation} \label{delta}
\hat{\delta}\equiv-\sin\delta(\cos\alpha\,\hat{i}+\sin\alpha\,\hat{j})+\cos\delta\,\hat{k}.
\end{equation}
Moreover, let $\hat{u}$  be a unit vector which lies in the orbital plane of the binary
along the line of nodes, which is defined to be the intersection of the orbital
plane with the tangent plane of the sky. Then $\hat{u}\cdot\hat{d}=0$ , and one can write
\begin{equation}
\hat{u}=\cos{\phi}\ \hat{\alpha}+\sin{\phi}\ \hat{\delta},
\end{equation}
where $\phi$ defines the  orientation of the line of nodes in the sky.
In the linearized theory of general relativity,  the metric
perturbation is given to lowest order by the second-time derivative of the quadrupole moment of the source (\citealt{Misner+etal+1973})
\begin{equation}\label{hab}
h^{\rm{TT}}_{ab}(t)=\frac{2}{d}\ddot{Q}_{ab}^{\rm{TT}}(t-d),
\end{equation}
where $d$ is the distance to the GW source, $h_{ab}^{\rm{TT}}$ describes the waveform of GWs in the transverse-traceless (TT) gauge, and $Q_{ab}^{\rm{TT}}$ is the quadrupole moment of the source evaluated in the retarded time $t-d$. For a GW travelling in a definite direction $\Omega$, the waveform of GWs is usually  written as
\begin{equation}
h_{ab}^{\rm{TT}}(t,\hat{\Omega})=h_{+}(t)\epsilon_{ab}^{+}(\hat{\Omega})
+h_\times(t)\epsilon_{ab}^\times(\hat{\Omega}),
\end{equation}
where $\hat{\Omega}=-\hat{d}$ is the unit vector pointing from the GW source to the SSB.
The polarization tensors are (\citealt{Lee+etal+2011}; \citealt{Ellis+etal+2012})
 \begin{equation}\label{eplus}
 \epsilon_{ab}^+(\hat{\Omega})=
 \left(\begin{array}{ccc}
 \sin^2\alpha-\cos^2\alpha\sin^2\delta & -\sin\alpha\cos\alpha(\sin^2\delta+1) & \cos\alpha\sin\delta\cos\delta\\
 -\sin\alpha\cos\alpha(\sin^2\delta+1) & \cos^2\alpha-\sin^2\alpha\sin^2\delta & \sin\alpha\sin\delta\cos\delta\\
\cos\alpha\sin\delta\cos\delta &  \sin\alpha\sin\delta\cos\delta & -\cos^2\delta\\
 \end{array}
\right),
\end{equation}
\begin{equation}\label{ecross}
 \epsilon_{ab}^\times(\hat{\Omega})=
 \left(\begin{array}{ccc}
 \sin(2\alpha)\sin\delta & -\cos(2\alpha)\sin\delta & -\sin\alpha\cos\delta\\
 -\cos(2\alpha)\sin\delta & -\sin(2\alpha)\sin\delta & \cos\alpha\cos\delta\\
 -\sin\alpha\cos\delta& \cos\alpha\cos\delta & 0\\
 \end{array}
\right),
 \end{equation}
For  a SMBHB
with a elliptical orbit, the polarization amplitudes of the emitting GWs are (\citealt{Wahlquist+1987})
\begin{equation}\label{hplus}
h_+(\theta)=H\{{\cos(2\phi)}[A_0+eA_1+e^2A_2]
-\sin(2\phi)[B_0+eB_1+e^2B_2]\},
\end{equation}
\begin{equation}
h_\times(\theta)=H\{{\sin(2\phi)}[A_0+eA_1+e^2A_2] \label{hcross}
+\cos(2\phi)[B_0+eB_1+e^2B_2]\},
\end{equation}
with $\phi$ being the orientation of the line of nodes, which is defined to be the intersection
of the orbital plane with the tangent plane of the sky. The parameters
in Eqs.(\ref{hplus}) and (\ref{hcross}) are
\begin{equation}
\begin{array}{lll}
H&\equiv&\frac{2^{8/3}\pi^{2/3}M_c^{5/3}(1+z)^{5/3}}{(1-e^2)P_{\rm{obs}}^{2/3}D_L},\\
A_0&=&-\frac{1}{2}[1+\cos^2(\iota)]\cos(2\theta),\\
B_0&=&-\cos(\iota)\sin(2\theta),\\
A_1&=&-\frac{1}{4}\sin^2(\iota)\cos\theta+\frac{1}{8}[1+\cos^2(\iota)][5\cos\theta+\cos(3\theta)],\\ B_1&=&\frac{1}{4}\cos(\iota)[5\sin(\theta)+\sin(3\theta)],\\
A_2&=&\frac{1}{4}\sin^2(\iota)-\frac{1}{4}[1+\cos^2(\iota)],\\\label{parameter}
B_2&=&0.
\end{array}
\end{equation}
Note that, compared to those shown in \citealt{Wahlquist+1987}, in Eq. (\ref{parameter})  we have chosen $\theta_n=0$, the value of $\theta$ at the line of nodes. Moreover, ${M_c}=\mu^{3/5}M^{2/5}$ is  the chirp mass with $\mu$ being the reduced mass of the binary system. The appearance of the factor $(1+z)^{5/3}$ included in the expression of $H$  is due to the effect of cosmological redshift.
   $\iota$ is the angle of inclination of the orbital plane to the tangent plane of the sky,
   $P_{\rm obs}=P(1+z)$ is the observational period of the binary, and $D_L$ is the luminosity distance from the binary system to the SSB.
   In the standard cosmology model, the luminosity distance is given by
\begin{equation}
D_L=\frac{1+z}{H_0}\int_0^z\frac{dz'}{\sqrt{\Omega_\Lambda+\Omega_m(1+z')^3}},
\end{equation}
where $H_0$ is the Hubble constant, $z$ is the cosmological redshift, and $\Omega_\Lambda$ and $\Omega_m$ are the density contrast of  dark energy  and  matter respectively.
From the observations of WMAP 9 (\citealt{Hinshaw+etal+2013}), one has $H_0=69.7$ km s$^{-1}$ Mpc$^{-1}$, $\Omega_\Lambda=0.72$ and $\Omega_m=0.28$.
For the particular case of $e=0$, Eqs. (\ref{hplus}) and (\ref{hcross}) reduces to (\citealt{Lee+etal+2011})
\begin{equation}
h_+(t)=h_0\left[\cos\iota\sin(2\phi)\sin(\omega_gt)
-\frac{1}{2}(1+\cos^2\iota)\cos(2\phi)\cos(\omega_gt)\right],\label{hcircular}
\end{equation}
\begin{equation}
h_\times(t)=-h_0\left[\cos\iota\cos(2\phi)\sin(\omega_gt)
+\frac{1}{2}(1+\cos^2\iota)\sin(2\phi)\cos(\omega_gt)\right],\label{hcircular2}
\end{equation}
where $h_0=2^{4/3}{M_c}^{5/3}\omega_g^{2/3}D_L^{-1}(1+z)$ with $\omega_g={4\pi}/{P}$ being the angular frequency   of the radiated  GWs at the source.
It is worth to point out that the time $t$ in Eqs. (\ref{hcircular}) and (\ref{hcircular2}) stands for the time scale
around the GW source. Alternatively, one can rewrites Eqs.(\ref{hcircular}) and (\ref{hcircular2}) using the time at observer $t'$. Due to the cosmological redshift effect, one has $t'=t(1+z)$ if the zero points of $t$ and $t'$ are chosen to be the same. One the other hand, the intrinsic frequency of the GWs will be suffered from a redshift, that is, $\omega_g^{(\rm{obs})}={\omega_g}/{(1+z)}$, where  $\omega_g^{(\rm{obs})}$ is the observational angular frequency of GWs.  Thus, one has $\omega_g t\equiv \omega_g^{(\rm{obs})}t'$, and $h_0$ can be expressed as $h_0=2^{4/3}{M_c}^{5/3}(\omega_g^{(\rm{obs})})^{2/3}D_L^{-1}(1+z)^{5/3}$.

\section{The pulsar timing residuals induced by a single GW}

The GW will cause a fractional shift in frequency, $\nu$, that can be defined by a redshift
 (\citealt{Demorest+etal+2013}; \citealt{Ellis+etal+2012}; \citealt{Anholm+etal+2009})
\begin{equation}\label{z}
z(t,\hat{\Omega})\equiv\frac{\delta\nu(t,\hat{\Omega})}{\nu}=-\frac{1}{2}\frac{\hat{n}^a\hat{n}^b}{1+\hat{n}\cdot
\hat{\Omega}}\epsilon_{ab}^{A}(\hat{\Omega})\Delta h_A(t),
\end{equation}
where
\begin{equation}
\Delta h_A(t)=h_A(t_e)-h_A(t_p).
\end{equation}
Here $A$ denotes ``$+,\times$'' and  the standard Einstein summing convention was used.  $t_e$
and $t_p$ are the time at which the GW passes the earth and pulsar, respectively.
Henceforth, we will drop the subscript ``$e$'' denoting the earth time unless otherwise noted.
 The unit vector, $\hat{n}$, pointing from the SSB to the pulsar is  explicitly written as
 \begin{equation}\label{n}
 \hat{n}=\cos{\delta_p}[\cos\alpha_p\hat{i}+\sin\alpha_p\hat{j}]+\sin{\delta_p}\hat{k},
 \end{equation}
 where    $\alpha_p$ and $\delta_p$ are the right ascension and declination of the pulsar, respectively.
From geometry on has (\citealt{Ellis+etal+2012}; \citealt{Anholm+etal+2009})
 \begin{equation}\label{tp}
 t_p=t-D_p(1-\cos\eta),
 \end{equation}
 where $D_p$ is the distance to the pulsar and $\cos\eta=-\hat{n}\cdot\hat{\Omega}=\hat{n}\cdot\hat{d}$ with
 $\eta$ being the angle between the pulsar direction and the GW source direction.
 Combining Eqs. (\ref{eplus})-(\ref{z}), we obtain
 \begin{equation}\label{redshift}
 z(t,\hat{\Omega})=-\frac{1}{2}(1+\cos\eta)\{\cos(2\lambda)
 [h_+(t)-h_+(t_p)]\\
 +\sin(2\lambda)[h_\times(t)-h_\times(t_p)]\},
  \end{equation}
 where $\cos\eta=\sin\delta_p\sin\delta+\cos{\delta_p}\cos{\delta}\cos(\alpha-\alpha_p)$, and
 $\lambda$ is defined as (\citealt{Wahlquist+1987})
 \begin{equation}
 \tan\lambda\equiv\frac{\hat{n}\cdot\hat{\delta}}{\hat{n}\cdot\hat{\alpha}}=
 \frac{\cos\delta_p\sin\delta\cos(\alpha-\alpha_p)-\sin\delta_p\cos\delta}{\cos\delta_p
 \sin(\alpha-\alpha_p)},
 \end{equation}
 where Eqs. (\ref{alpha}), (\ref{delta}) and (\ref{n}) were used. It can be found from Eq. (\ref{redshift}) that,
 the frequency of the pulses from the pulsar will suffer no shift from the GW
 for $\eta=0^\circ$ and $\eta=180^\circ$ allowing for Eq. (\ref{tp}).

 The pulsar timing residuals induced by GWs can be computed by integrating the redshift given
 in Eq. (\ref{redshift}) over the observer's local time (\citealt{Hobbs+etal+2009}; \citealt{Lee+etal+2011}; \citealt{Ellis+etal+2012}; \citealt{Anholm+etal+2009}):
 \begin{equation}
 R(t,\hat{\Omega})=\int_0^tz(t',\hat{\Omega})dt'.
 \end{equation}
 For the particular case of $e=0$, one has the analytic expressions of the timing residuals as follows,
 \begin{equation}\nonumber
 \begin{array}{lll}
 R(t,\hat{\Omega})&=&\frac{h_0\sin(\Delta\Phi/2)}{2\omega_g(1-\cos\eta)}
 \{[C_+\cos(2\phi)+C_\times\sin(2\phi)](1+\cos^2\iota)\cos(\omega_gt-\Delta\Phi/2)\\
 & &+2[C_\times\cos(2\phi)-C_+\sin(2\phi)]\cos\iota\sin(\omega_gt-\Delta\Phi/2)\},
 \end{array}
 \end{equation}
 where $C_+$, $C_\times$ and $\Delta\Phi$ depend on the geometrical configuration of the pulsar and GW source  by
 \begin{equation}\nonumber
 \begin{array}{lll}
 C_+&=&\frac{1}{4}\cos^2\delta_p\{2\cos^2\delta+[\cos(2\delta)-3]\cos[2(\alpha_p-\alpha)]\}
 -\cos^2\delta\sin^2\delta_p\\
 & &+\cos\delta_p\cos(\alpha_p-\alpha)\sin\delta_p\sin(2\delta)
 \end{array}
  \end{equation}
 \begin{equation}
 C_\times=\cos\delta\sin(2\delta_p)\sin(\alpha_p-\alpha)-\sin\delta\cos^2\delta_p\sin[2(\alpha_p-\alpha)],
 \end{equation}
 \begin{equation}
\Delta\Phi=\omega_gD_p(1-\cos\eta).
\end{equation}
  It is also interesting to calculate the standard deviation of the time residuals for a definite direction of GWs, $\sigma_R$, which is defined as (\citealt{Hobbs+etal+2009})
\begin{equation}\label{sigmar}
\sigma_R=\left[\frac{1}{T}\int_0^T R^2(t)dt-\left(\frac{1}{T}\int_0^T R(t)dt\right)^2\right]^{1/2},
\end{equation}
where $T$ can be chosen as the period $P$  due to the periodicity of $R(t)$.

\section{The properties of the  timing residuals affected by all the related parameters}

As can be seen from the above, the pulsar timing residuals are related to  many parameters,
such as $e$, $\phi$, $\iota$, $\lambda$, and $\eta$. The different properties of $R(t)$ and $\sigma_R$ due to different values of $e$ with a constant $H$ have been studied preliminarily in \citealt{Tong+etal+2013a,Tong+etal+2013b}.
In this section, we analyze the properties of $R(t)$ and $\sigma_R$ induced by all the related parameters.  SMBHBs are the ideal GW sources detected by pulsar timing
arrays (\citealt{Hobbs+etal+2009}; \citealt{Sesana+Vecchio+2010}), whose response frequencies are in the range $\sim10^{-9}-10^{-7}$ Hz. Hence we can set the observed orbital period of  a SMBHB to be $P_{\rm{obs}}=10^9$ s for instance, since the frequencies of the GWs from a SMBHB
with an elliptical orbit are a few or tens times of the orbital frequency of the binary system (\citealt{Wahlquist+1987}; \citealt{Maggiore+2008}).
With fixed values of all the  related parameters, one can calculate $R(t)$ and $\sigma_R$ respectively. As shown in \citealt{Tong+etal+2013a}, $R(t)$ presents quite different for
different $e$. Thus, one can infer that $R(t)$ will  presents complexities for different  values of various parameters. We will not illustrate the properties of $R(t)$ in this paper but
 instead of focusing on the more essential $\sigma_R$. Below, for illustration,  we assume $H=10^{-15}$, which is below the upper limit given in \citealt{Yardley+etal+2010},
 and take $e=0.3$ for instance, except additional analysis on the different values of $e$.

\subsection{The effects of $\lambda$ and  $\eta$}

First of all, we fix $\phi=\iota=0$ following \citealt{Wahlquist+1987} and \citealt{Tong+etal+2013a}. The waveforms of
 the polarization amplitudes $h_+(t)$ and $h_\times(t)$ for different $e$ were shown in \citealt{Wahlquist+1987}. Moreover, we always set $D_p=157$ pc, the distance from the SSB to PSR J0437-4715 (\citealt{Verbiest+etal+2008}). Thus, there are only two free parameters, $\lambda$ and $\eta$. Firstly,
 set $\eta=60^\circ$, we get $\sigma_R=33$ ns for all the values of $\lambda\in[0,360^\circ]$.
That is, $\sigma_R$ is nothing to do with $\lambda$. Therefore, we will set $\lambda=0$ in the following without losing generality. Secondly, we discuss the influences of $\eta$. In
Fig.1, we plot $\sigma_R$ versus $\eta$. One can see that, $\sigma_R$ has a decreasing trend  with $\eta$ and decays to be zero at $\eta=180^\circ$. Therefore, if the pulsar and the GW source
are almost in the same direction to the observer, the pulsar signals will be
 affected by GWs distinctly. However, one should note
that $\sigma_R=0$ for $\eta=0$ due to Eq. (\ref{tp}).

\begin{figure}
   \centering
   \includegraphics[width=14.0cm, angle=0]{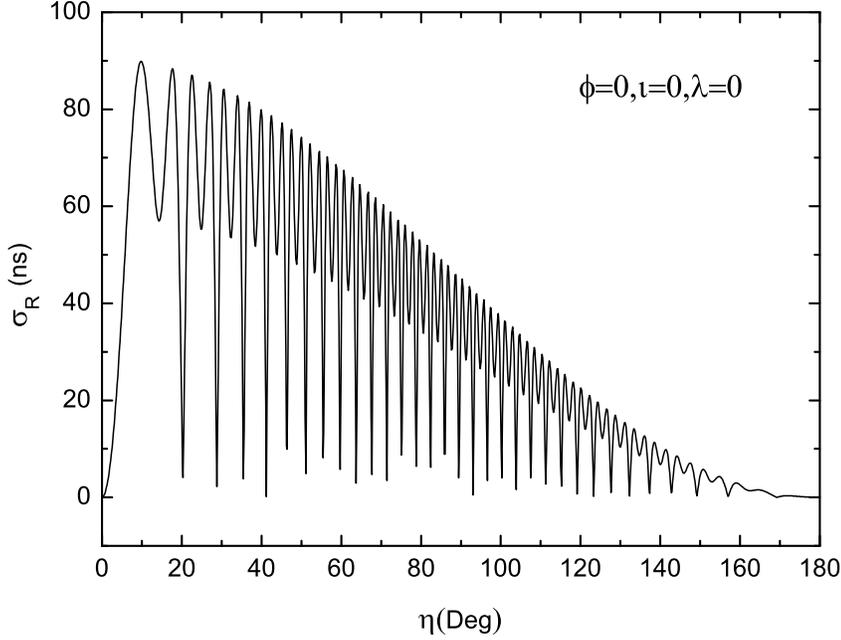}
   \caption{ The variety  of $\sigma_R$ along with $\eta$ for $e=0.3$ and $H=10^{-15}$. }
   \label{Fig1}
   \end{figure}

\subsection{The effects of $\phi$ and $\iota$}

For a full analysis, now we discuss  the influences of different $\phi$ and $\iota$ on $\sigma_R$. As can be seen in Eqs. (\ref{hplus}) and (\ref{parameter}), different values
of $\phi$ and $\iota$ will change the waveforms of $h_+(t)$ and $h_\times(t)$. Fig.2 shows $h_+(t)$ and $h_\times(t)$ for $\phi=0$ and $\phi=45^\circ$, respectively, for a fixed $\iota=60^\circ$. Similarly, Fig.3 shows $h_+(t)$ and $h_\times(t)$ for $\iota=0$ and $\iota=90^\circ$, respectively, for a fixed $\phi=30^\circ$.

\begin{figure}
   \centering
   \includegraphics[width=10.0cm, angle=0]{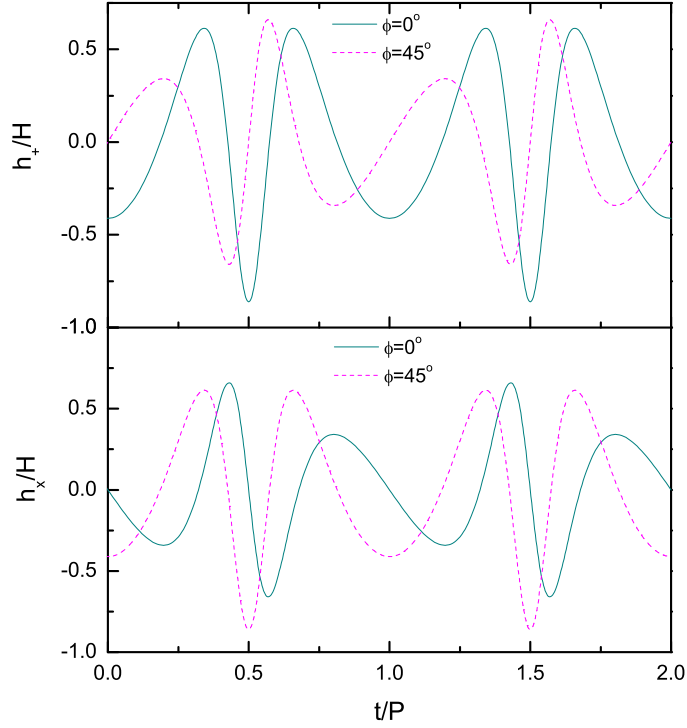}
   \caption{ $h_+(t)$ and $h_\times(t)$ normalized by $H$ for  $\phi=0$ and $\phi=45^\circ$,
respectively, with $e=0.3$ and $\iota=60^\circ$. }
   \label{Fig2}
   \end{figure}

\begin{figure}
   \centering
   \includegraphics[width=10.0cm, angle=0]{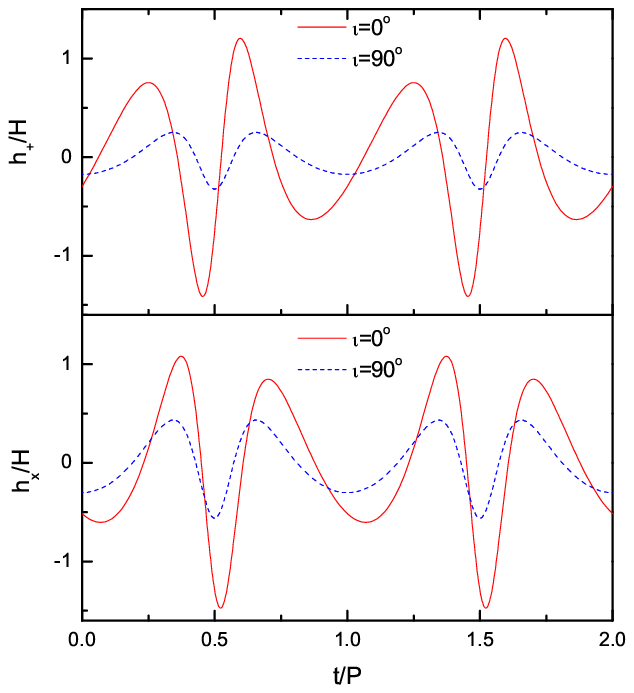}
   \caption{$h_+(t)$ and $h_\times(t)$ normalized by $H$ for  $\iota=0$ and $\iota=90^\circ$,
respectively, with $e=0.3$ and $\phi=30^\circ$.}
   \label{Fig3}
   \end{figure}

 Using the concrete forms of $h_+(t)$ and $h_\times(t)$, we can calculate the corresponding $\sigma_R$
 with the help of  Eq. (\ref{sigmar}).  Setting $\eta=60^\circ$, the resulting $\sigma_R$ versus $\phi$ with fixed $\iota=60^\circ$ is shown in Fig.4. Similarly, the resulting $\sigma_R$ versus $\iota$ with fixed $\phi=30^\circ$ is shown in Fig.5.
 From Fig.4, it can found that the maximal change of $\sigma_R$ is about $4.1$ ns due to different $\phi$. On the other hand, the maximal change of $\sigma_R$  due to different $\iota$ is about $24.7$ ns as shown in Fig.5. Therefore, $\iota$ affects $\sigma_R$ more distinctly than $\phi$ does. This is implied in Fig.2 and Fig.3,
 which show that $\iota$ affects the polarization amplitudes of GWs more distinctly than $\phi$ does.
 Note that,  $\sigma_R$ is proportional to the the polarization amplitudes of GWs, and in turn,  $H$.
 Moreover, $\sigma_R$ presents periodicity along with $\phi$  with a  period  being $\pi/2$,
 and $\sigma_R$ is symmetrical relative to $\iota=90^\circ$ where $\sigma_R$ has a minimal value. Hence, the SMBHB with an orbital plane perpendicular to the line of sight of the observer will
 contribute to the timing residuals least.

\begin{figure}
   \centering
   \includegraphics[width=10.0cm, angle=0]{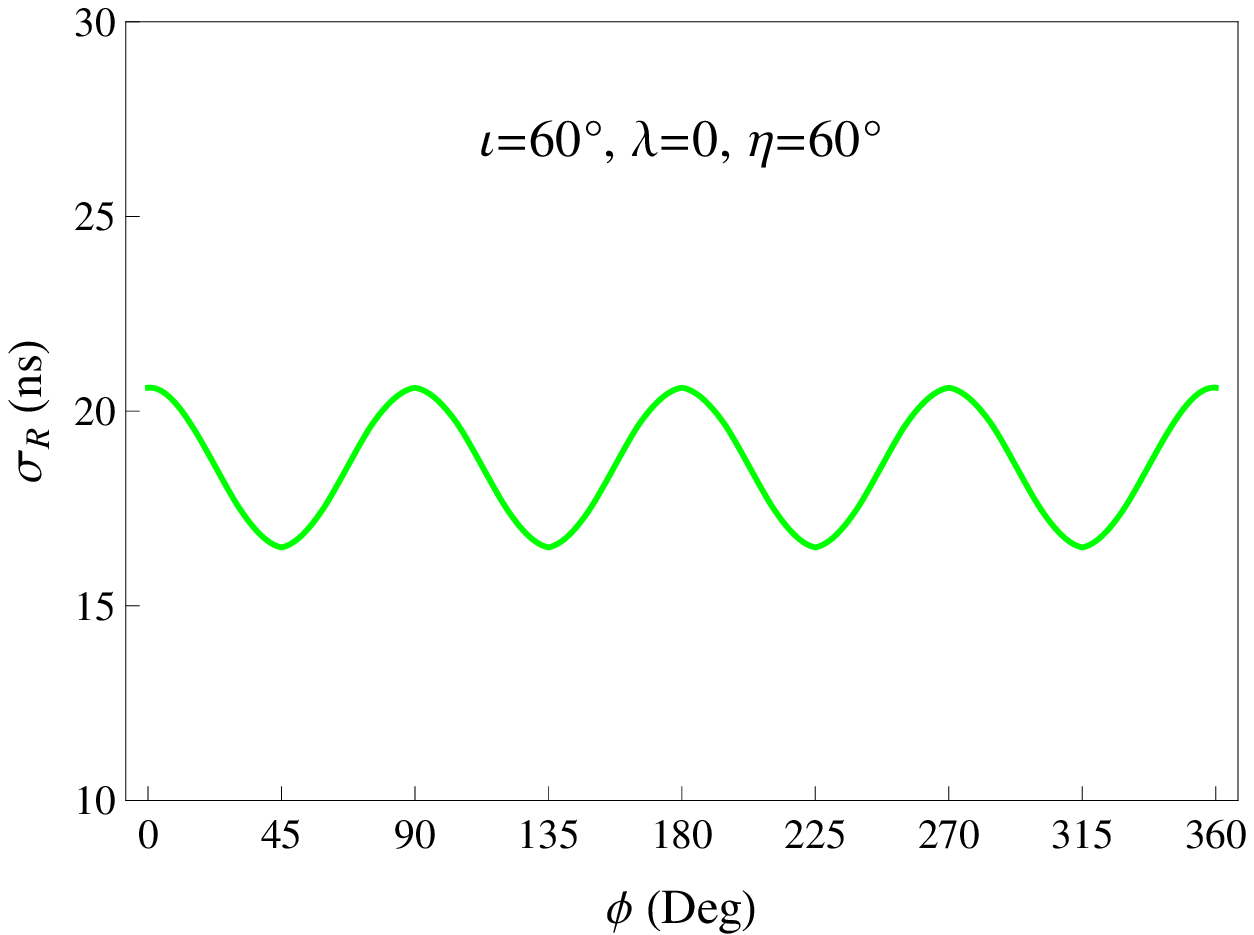}
   \caption{ The variety  of $\sigma_R$ along with $\phi$ for $e=0.3$.}
   \label{Fig4}
   \end{figure}

\begin{figure}
   \centering
   \includegraphics[width=10.0cm, angle=0]{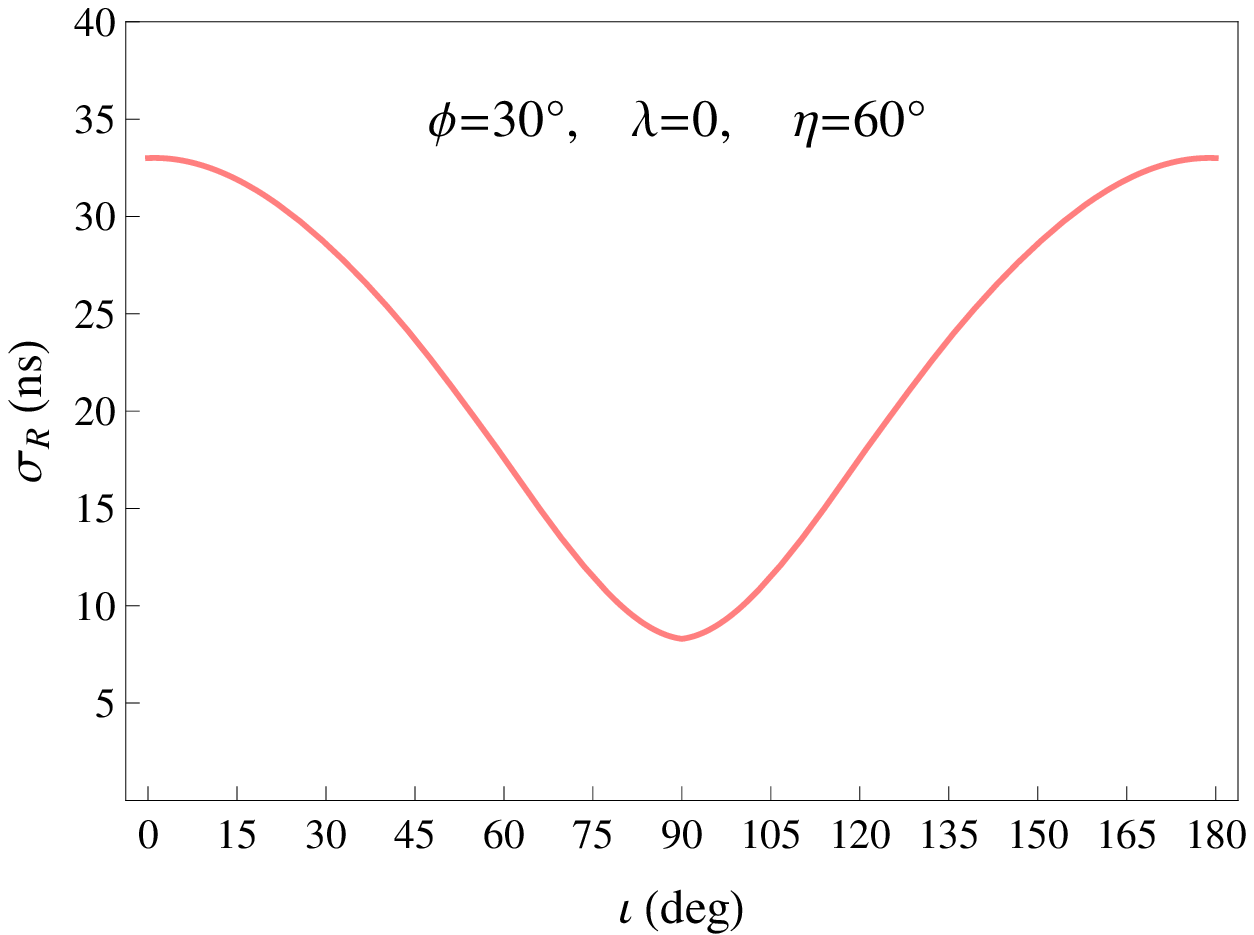}
   \caption{The variety  of $\sigma_R$ along with $\iota$ for $e=0.3$.}
   \label{Fig5}
   \end{figure}

\subsection{The effect of $e$}

The effect of different $e$ on the timing residuals has been studied in \citealt{Tong+etal+2013a}, for a fixed
value of $H$. Here,
we focus on the effect of $e$ on the standard deviation $\sigma_R$. First of all, the waveforms of the polarization amplitudes are quite different for different $e$ with $\phi=\iota=0$, as shown in \citealt{Wahlquist+1987}.  When calculating $\sigma_R$, we consider two cases with different values of $H$. Firstly, we fix $H=10^{-15}$. Secondly, we fix $H'\equiv H|_{e=0}=\frac{2^{8/3}\pi^{2/3}M_c^{5/3}(1+z)^{5/3}}{P_{\rm{obs}}^{2/3}D_L}=10^{-15}$.
For a concrete set of parameters $\phi=30^\circ$, $\iota=60^\circ$, $\lambda=0$ and $\eta=45^\circ$, the resulting $\sigma_R$ along with $e$ for the two cases are listed  in Table 1 and Table 2, respectively.
  Furthermore,  interpolation curves of the data in Table 1 and Table 2
are  plotted together in Fig.6.
 One can see clearly that, for a definite set of parameters, $\sigma_R$ decreases distinctly with  larger values of $e$  in the case of $H=10^{-15}$, however, $\sigma_R$ changes relative small and levels off for $e\geq0.6$ in the
 case of $H'=10^{-15}$. The two cases give rise to quite different results especially for larger values of $e$. Therefore, when doing simulations of the GW sources, one should clarify the assumptions since the discrepancy exists between the two cases.

\begin{table}
\bc
\begin{minipage}[]{100mm}
\caption[]{$\sigma_R$ for different $e$ with $H=10^{-15}$. The set of parameters were chosen as
 $\phi=30^\circ$, $\iota=60^\circ$, $\lambda=0$ and $\eta=45^\circ$.\label{tab1}}\end{minipage}
\setlength{\tabcolsep}{10pt}
\small
 \begin{tabular}{c|cccccccccc}
  \hline\noalign{\smallskip}
$e$ & $0$ & $0.1$ & $0.2$& $0.3$ &$0.4$ &$0.5$ & $0.6$ & $0.7$ & $0.8$ & $0.9$  \\
\hline
$\sigma_R$ (ns) & $50.8$ &49.6 &46.2 &41.1 & 35.3 & 29.6 & 24.4 & 19.3 &  13.8 & 7.3  \\
\noalign{\smallskip}\hline
\end{tabular}
\ec
\end{table}

\begin{table}
\bc
\begin{minipage}[]{100mm}
\caption[]{$\sigma_R$ for different $e$ with $H'=10^{-15}$.
The set of parameters were chosen exactly the same as Table 1.\label{tab2}}\end{minipage}
\setlength{\tabcolsep}{10pt}
\small
 \begin{tabular}{c|cccccccccc}
  \hline\noalign{\smallskip}
$e$ & $0$ & $0.1$ & $0.2$& $0.3$ &$0.4$ &$0.5$ & $0.6$ & $0.7$ & $0.8$ & $0.9$  \\
\hline
$\sigma_R$ (ns) & $50.8$ &50.1 &48.1 &45.2 & 42.0 & 39.5 & 38.1 & 37.8 &  38.3 & 38.4  \\
\noalign{\smallskip}\hline
\end{tabular}
\ec
\end{table}

\begin{figure}
   \centering
   \includegraphics[width=10.0cm, angle=0]{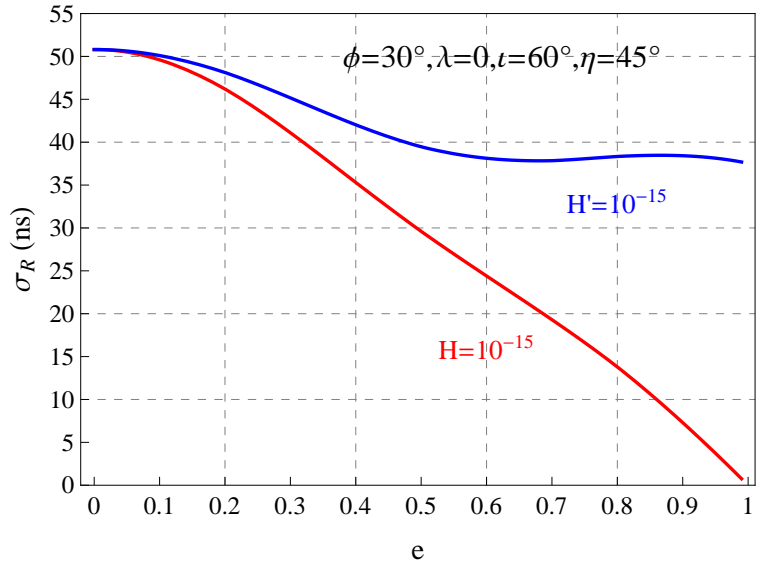}
   \caption{The varieties  of $\sigma_R$
    along with $e$ for $H=10^{-15}$ and $H'=10^{-15}$, respectively.}
   \label{Fig6}
   \end{figure}

\section{Conclusions and discussions}

The pulsar timing residuals induced by GWs from single sources are related to many parameters.
We have analyzed the effects of various parameters on the timing residuals $R(t)$ and their standard deviations $\sigma_R$. Among all the related parameters, different  $\lambda$ will not change $\sigma_R$. Generally speaking, a larger $\eta$ leads to a smaller $\sigma_R$ except $\eta=0$.
On the other hand, $\sigma_R$ presents periodicity along with $\phi$ with a period equalling $\pi/2$,
and $\sigma_R$ is symmetrical relative to $\iota=90^\circ$, where $\sigma_R$ has a minimum.  Besides,
For definite other parameters, $\sigma_R$ decreases with larger $e$ in the case of fixed value $H=10^{-15}$, however, $\sigma_R$ will not change so much in the case of $H'\equiv H|_{e=0}=10^{-15}$.
It is worth to note that the timing residuals and the standard deviations are both proportional to the polarization amplitudes, and in turn $H$. By comparison, the parameters $\eta$, $\iota$ and $e$ affect $\sigma_R$ evidently. If one want to detect single GW sources, under the lack of essential information about the sources, the sensitivities of these parameters analyzed above could give a hint on how to reduce the space parameters. On the other hand, for the study of the pulsar timing standard, the GWs is one kind of timing noise. Knowing how a strong GW source like the SMBHB in the blazar OJ287 (\citealt{Sillanpaa+etal+1996}) affect the timing signals from pulsars, is very important in constructing pulsar timing standard.

 Note that, all the results are based on the non-evolving single sources. The evolving sources especially the SMBHBs at the merge phase will radiated much stronger GWs, which are easier to be detected. So the
pulsar timing residuals induced by GWs from the merge of SMBHBs are worth to be studied elsewhere.
In this case, the single GW source is instantaneous other than continuous, and the eccentricity will
decay almost to be zero.

\normalem
\begin{acknowledgements}
This work was supported by National
Natural Science Foundation of China (Grant No 11103024 and 11373028) and
  the  program of the light in China's Western Region of CAS.
\end{acknowledgements}

\bibliographystyle{raa}
\bibliography{bibtex}

\end{document}